# ATLAS Data Challenge 1

G. Poulard
*CERN, Geneva, SWITZERLAND*
On behalf of the ATLAS Data Challenge team.

The ATLAS Collaboration at CERN is preparing for the future data taking and analysis at LHC that will start in 2007. To validate its Computing Model, its complete software suite, its data model, and to ensure the correctness of the technical choices it has been decided to run series of so-called Data Challenges. In 2002 the main goals of these Data Challenges were the preparation and the deployment of the full Monte Carlo software chain and the setting up of the computing infrastructure in view of a worldwide production of large event samples. The ATLAS High Level Trigger community that has to prepare a Technical Design Report by mid-2003 mostly requested these data. It should be noted that it was not an option to "run everything at CERN" even if we wanted to; the resources are not available at CERN to carry out the production on a reasonable time-scale. We have therefore had to face the great challenge of organizing and then carrying out this large-scale production at a significant number of sites around the world. However, the benefits of this are manifold: apart from realizing the required computing resources, this exercise builds worldwide momentum for ATLAS computing as a whole. The biggest part of the production has been run in the conventional way but a significant part was also run in few GRID testbeds.

## 1. INTRODUCTION

The LHC Computing Review [1] recommended having data challenges (DC) of increasing size and complexity. The ATLAS collaboration [2] planned to run several of these DCs starting with an initial one called DC0 in spring 2001 followed by DC1 which had a main goal to provide simulated and reconstructed data for High Level Trigger and physics studies. DC2 will follow in 2004 having as a goal to provide input for the Computing Technical Design Report due by mid 2005. Further DCs will focus on the readiness of the full ATLAS software suite and infrastructure for the commissioning of the ATLAS apparatus in 2006 and the initial data taking in 2007.

These DC exercises will be performed in the context of the LHC Computing Grid (LCG) project [3]. The experience gained will be used not only to formulate the ATLAS Computing TDR but also to help the formulation of the LCG TDR.

LCG is a deployment of the Grid technologies for the LHC computing. The Grid technologies promise several advantages for a multinational, geographically distributed project: they allow for a uniform infrastructure of the project computing-wise, simplify the management and coordination of the resources while potentially decentralizing such tasks as software development and analysis, and last, but not least, the Grid is an affordable way to increase the computing power. If the ATLAS Data Challenges will demonstrate that usage of the Grid, indeed, gives all those advantages, the collaboration should become committed to "gridification" of its sites and tools, by making use of the best available Grid middleware.

## 2. ATLAS DATA CHALLENGES

During the LHC preparation phase, all experiments have large needs for simulated data, to design and optimise the detectors. This "Monte Carlo" simulation is done in the following steps:

- Particles emerging from the collisions (called collision final state or simply final state) are generated using programs usually based on physics theories and phenomenology (called generators);
- The particles of the generated final state are transported through the virtual detector according to the known physics laws governing the passage of particles through matter;
- The resulting interactions with the sensitive elements of the detector are converted into rates of electronic counters (digitisation) similar to those produced by the real detector;
- The events are reconstructed.

The (Monte Carlo) generated information (sometimes called *truth*) is saved for comparison with the reconstructed information.

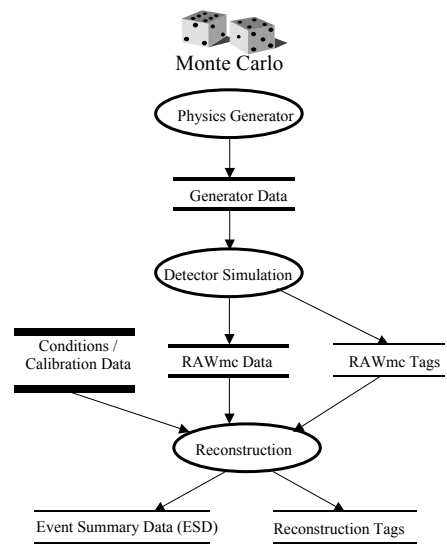

## 2.1.  DC0

The initial step of the ATLAS Data Challenges, called DC0, was decided in 2001 and had a first scope of preparing the software suite and the computing infrastructure for the next DCs.

DC0 comprises a 'continuity' test through the software chain. The aim was primarily to check the state of readiness for DC1.  It was intended to simulate and reconstruct samples of the order of 100k "Z+jet" events, or similar. The number of events was less important than checking that the software works; the primary goal was to ensure that the full software chain was ready. Issues to be checked include:

> Geant3 simulation on PC farm
> 'pile-up' handling
> reconstruction running

In addition few samples of single particle data (electrons, muons, photons,) were generated to finalise the study of the impact of the new detector layout.

DC0 ran at a single centre; however, the major centres intending to participate in DC1 were encouraged to perform some generations and to test the software. Note that, assuming a typical PC can simulate about 20 events an hour, then a farm of 100 PCs can generate 100k events in a few days. So the hardware (CPU and storage) was not a big issue.

## 2.2.  DC1

DC1 was scheduled to run from April 2002 to early part of 2003. As stated earlier the main goal was to provide simulated data to the High Level Trigger (HLT) community that has to prepare a Technical Design Report (TDR) by mid-2003. It was divided into three phases. In the first phase, April-August 2002, we put in place the infrastructure and the production tools  to be able to run the 'massive' production worldwide. The second phase, started in October 2002, the goal was to produce the pile-up data. The third phase was the reconstruction of the simulated data, since it had to include algorithms from the HLT it was not considered, as fully be part of DC1. The reconstruction has been done with the 'pure' offline reconstruction code the addition of the HLT part is expected in the next coming weeks.

The event generation can use several event generators (e.g. Pythia, Herwig, Isajet, etc.) and can run either in the Fortran Atlsim framework or in the official ATLAS Athena framework. The fast simulation, Atlfast, was used to control the quality of the generated data.

The current detector simulation code called DICE is Fortran based, uses Geant 3.21 to track the events through the detector and runs in the Atlsim framework. Events are written out in the form of Zebra banks.

Most of the reconstruction programs have now moved to OO, even if some packages are still in Fortran. The new reconstruction uses the Athena framework and in the current situation input data can come from ZEBRA [4] or ROOT-IO [5].

## 3.  GENERATION AND SIMULATION

### 3.1.  Event Generation

The generation of all event samples was done at CERN using Pythia 6.203 running inside Athena. The events were converted into HepMC [6] and then written out into ROOT I/O using the Athena-Root conversion service.

Several samples of physics events were generated. Among them:  **"jet"; "minimum bias"; single W; single Z**; **W+jet; Z+jet; Photon+jet; inclusive top; Higgs** and **MSSM Higgs Samples;** with different characteristics (transverse momentum; decays, etc).

### 3.2.  Generation Monitoring

The quality of the generated events produced was assessed and controlled with the use of histograms of various characteristic properties of those events. These histograms were produced after the generation by running an independent process that invoked a purpose-written algorithm called HistSample and written to an RZ-format output files. An n-tuple was also written into the same file.

This n-tuple contains quantities related to the jet structure of the event. Jet finding was performed by running Atlfast with the normal smearing turned-off, and then making use of the associated Atlfast utilities to perform the jet finding at the particle level in the generator output. The n-tuple was then used in a secondary job, which runs a KUMAC in the PAW [7] environment to produce histograms of the number of reconstructed jets, their $p_T$ spectra and pseudo-rapidity distributions. These are normalized in various ways: to the number of events; to the number of jets; and to the total cross section.

Finally, the two-histogram samples were merged and a postscript summary of all of the histograms produced was made and checked for consistency with the physics expectations for the given sample. The various output files were put into long-term storage using the CERN Advanced Storage Manager CASTOR [8].

### 3.3.  Event Simulation

The ATLAS detector simulation was done in the Atlsim framework using Geant3. Atlsim is a PAW-based framework, which uses KUIP [9] for job control. It has an improved memory management, eliminating any hard limits on the track/vertex/hit numbers. It also has improved hadronic physics based mainly on the GCALOR package. To avoid known problems with low energy $K^0_L$ (zero cross section in FLUKA), they are always traced by GEISHA.

Atlsim uses plug-in components (shared libraries) to provide extra I/O facility (e.g. ROOT) and to load ATLAS detector geometry. Description of the ATLAS geometry is taken from the DICE package.

During the simulation-phase di-jet events produced by Pythia where analyzed by a filtering routine which looked for a predefined energy deposition in two neighboring towers in η-φ space. Only events selected by the filter were passed to the simulation step and then written out.

## 3.4. Quality assurance and data validation

The aim of the ATLAS DC validation [10] has been to insure the compatibility and reproducibility of the samples produced by different simulation sites. In addition, the validation process has served as a general monitoring tool of changes/improvements to the ATLAS detector geometry. A histogram-by-histogram comparison is performed between two sets of validation histograms, providing a bin-by-bin significance plot and a chi2. To accomplish this task a semi-automated system to compare and identify differences between two simulations was set-up. The validation test-suite consists of a modular analysis structure based on PAW, which runs off a general-purpose n-tuple (CBNT) from the ATLAS reconstruction framework (ATRECON), with information on MC event generation and reconstruction by all ATLAS sub-detectors.

The analysis procedure consists of two steps. First, a (open-ended) list of sub-detector specific macros is run from a master process to produce the validation histograms for the comparison. Thereafter, a histogram-by-histogram comparison is performed between two sets of validation histograms, providing a bin-by-bin significance plot and a chi2. At the end a summary chi2-bar chart for all compared histograms is made.

A broad variety of validation samples of dedicated single particle scans and physics benchmark processes (H, Z) were produced to validate the full simulation chain. This initial phase proved to be important for the discovery of missing/faulty/new aspects of the detector geometry implemented in the simulation. This phase demonstrated also, that for the validation of a larger set of samples, the quality and stability of services such as AFS, CASTOR or batch system have to be very good in order to reduce the amount of effort and time needed.

The validation of participating institutions was done by comparing the simulation of identical input samples from different sites and by comparisons of larger, statistically independent, samples of the same physics process. The validation provided an important checking of the simulation infrastructure at the contributing DC sites. During the initial phase it was a very complex and intensive but absolutely necessary activity.

The physics validation of the DC1 data was done in parallel. A di-jet sample (15M events) was processed with the fast simulation and reconstruction package, Atlfast. A comparison of the events content for multiplicities of jets, b-jets, c-jets, electrons and photons, with a similar sample produced in '96/97 large-scale production, was performed. In addition fully simulated samples have been inspected and are presently used e.g. for detailed detector calibration purposes.

## 4. PILE-UP

### 4.1. Pile-Up Procedure

Any collision registered in the ATLAS detector, contains in fact a superposition of particles coming from a single "physics" event, which triggered the readout, and of particles coming from another un-selected pp collisions.

On average per every bunch crossing seen by the ATLAS detector or a subsystem one expects to have 23 "unbiased" overlaying collisions. The total number of observed particle per event depends on the signal collection time, which varies from few ns in silicon detectors to about 700 ns in the Muon Drift Tubes (MDT).

In addition, while in Liquid Argon (LAr) calorimeters the signal is measured shortly after the trigger so that it is affected only by previous bunch crossings, measurements in drift detector such as Transition Radiation Tracker (TRT) or MDT continues for the maximum signal collection time so that they are sensitive to the same amount of posterior bunch crossings.

The full pile-up is simulated as a number of minimum bias collisions properly distributed in time and overlaying the physics collision.

As every collision is normally simulated only for few 100 ns of the propagation time, there is one additional component missing in this scheme: Neutrons may fly in the ATLAS cavern for few second until they are thermalised, thus producing kind of a permanent neutron-photon gas which creates a constant rate of Compton electron and spalation protons observed in the muon system. This component, i.e. additional hits created by long living particles, is called "cavern background".

Cavern background is simulated as a separate component that is added on top of every single minimum bias event. This is done in the following steps:

- 1.) A standalone dedicated Geant3/GCALOR based detector simulation program with improved neutron propagation and a simplified ATLAS geometry is run on pp collisions. The output of this program provides particle fluxes in the envelopes surrounding muon chambers. The fluxes are provided as list of particles with all related parameters per a pp interaction on the entrance of each chamber envelope.

- 2.) Atlsim randomly reads from these fluxes an average number of particles per single pp collision and feeds a subset of them into ATLAS DICE geometry. At this moment all photons and neutrons entering the chamber envelopes are selected ($E_{kin}$>10 KeV). Charge particles are selected only the first time they appear in the output list and only if their production time is bigger than the time cut-of of the DICE simulation, so that the prompt component of the calorimeter punch-through is not double counted. The starting time of all selected particles is reset to 0-25 ns interval.

A significant randomization is achieved at this moment due to:
- random initial particle selection;
- low probability of neutron and photon interaction in the chamber envelopes;
-   arbitrary selected particle rotation at the input

This allows multiple re-use of the particle fluxes simulated in the first, the most CPU consuming   step.

The detailed muon system geometry description provided by DICE is used to simulate signals induced by the cavern particles in the muon chambers.

The initially selected neutral particles are propagated only within chamber envelopes to avoid double counting of the n-gamma cascade. However, all their products and initially selected charged particles are trace until the GEANT program stops them.

Hits produced during the tracking (usually in the same 0-25 ns time range) are saved in pseudo-events normalized per one pp collisions as a standard Atlsim) simulation output.

- 3.) Output of the cavern background simulations is mixed with the standard fully simulated minimum bias events, thus producing new minimum bias events with the cavern backgrounds included. Mixing proportion may vary from 1 to 10 as the "safety factor". ($K^0$ and their decay product are already correctly simulated to some extend in the normal minimum-bias tapes as Atlsim contains the known bug correction for the $K^0$ propagation) This approach drastically reduces the time need to simulate the signals induced in the muon spectrometer by the cavern background comparing to the previously used technique. In the same time it allows for a realistic Compton electron and spallation proton production, which takes into account, all geometry details available in DICE properly convoluted with dedicated the n-gamma fluxes calculations.

- 4.) The resulting minimum-bias events should be added as a pile-up to any physics events. This should be done taken into account the LHC luminosity and bunch structure. To fully simulate the complete detector pile-up mixing should be done for +/- 30 bunch crossings with the average number varying from 4.6 events per bunch crossing for the low luminosity (L=2*10^{33}) run to 23 events per bunch crossing for the high luminosity (L=10^{34}) run.

CPU and memory requirements:

Step (1) is made once for a muon system layout. We need about 10K simulated events, which is only a small fraction of regular flux calculations.

Step (2) is also done once by a special version of Atlsim with the standard DICE geometry. This step takes about 6 SI95 second (SI95-s) sec per simulated event and requires standard Atlsim memory (<100 MB per job). The output is produced in files that contain 10K event. This is more than is needed for one to one file mixing at any reasonable safety factor. The total number of events needed at this step is about 10 Million (1000 files of 10K events each), the simulation time of the order of 60*10**6 SI95-s.

Step (3) should be done several times per each "minimum bias" tape (for every selected safety factor 1,2,5 as planned for the moment). As each job requires one "minimum bias" input file and one "cavern background" file, all three mixing could be done in one job. Each such job requires less than 2000 SI95-s but is output extensive (each 300MB input file yields 3 files close to one GB in total).

All together 1000 pre-mixing jobs are needed. The resulting files should be distributed over the production sites involved in the physics pileup production.

Step (4) is the most time consuming procedure as in addition to the event mixing it requires running full digitisation of the ATLAS detector. Time require per job does not depend on physics but on the luminosity only. A high luminosity pile-up job requires a 500 MB machine and takes 4400 SI95-s (800 for mixing and 3600 for digitisation).

This step produces output events of about ~<8 MB at high luminosity independent on the input physics event size.

The memory requirement (500 MB) was a matter of concern at the beginning of the exercise it was improved with more recent releases.

## 4.2. Resources needed for Pile-up production

As for the standard Atlsim/Geant3 simulation the digitization is accounted for in the simulation. We estimate the following average numbers for piled-up events in the $\eta$ range $|\eta| < 3$:

| Luminosity | Output size/event (MB) | CPU-time/event (SI95sec) |
|---|---|---|
| $2*10^{33} cm^{-2} s^{-1}$ | 3.6 | 2000 |
| $10^{34} cm^{-2} s^{-1}$ | 7.5 | 8000 |

## 5. RECONSTRUCTION

The standard ATLAS reconstruction chain was used. For HLT studies we concentrated on the algorithms for the Inner Detector Tracking system and the electromagnetic and hadronic calorimeters. The typical numbers for reconstruction are the following:

| Luminosity | Output size/event (MB) | CPU-time/event (SI95sec) |
|---|---|---|
| $2*10^{33} cm^{-2} s^{-1}$ | 0.02 | 3000 |
| $10^{34} cm^{-2} s^{-1}$ | 0.03 | 7600 |

As mentioned earlier a new pass over the data is in preparation, waiting for the dedicated Level 1 and Level 2 trigger algorithms.

## 6. BOOKKEEPING AND DATABASES

Essential components required for ATLAS Monte Carlo production are the associated bookkeeping and meta-data services. For DC1 the ATLAS Metadata base Interface, AMI [11], and the Replica Catalogue, Magda [12], were used to store information such as the logical and physical filenames, as well as a detailed description of the datasets.

## 6.1. The AMI Database

The term "bookkeeping" is used in many different contexts in HEP. Here we use the term to refer to a database application whose purpose is to store data that describes binary physics data that may be either real detector output, or, as in the case of DC1, simulated Monte Carlo data. The application is more precisely described as an "Application Metadata Catalogue". The term "Application" signifies that the information that is stored is specific to the application, and could not be guessed by some outside system, for example Grid

software. The aim of the Application Metadata catalogue is twofold:
- to make it possible to understand the contents of a file without actually having to open it,
- to search for a data filename or list of filenames, given a set of attributes of data.

The bookkeeping application used in DC1 is part of a development at the LPSC Grenoble. It was first used as an online electronic notebook for LAr test-beam data, and later adapted at the LPSC as a prototype application to store the metadata of offline calculations. This prototype was constantly upgraded during the running of DC0 and DC1. A presentation (MONT003) has been given in this conference.

### 6.1.1. Database Design

The bookkeeping application uses a layered architecture. It is written in JAVA, and makes use of the generic JDBC library for SQL database communication. In consequence, it is independent of platform, operating system and database technology. The only prerequisite is that java is installed on the client system. The architecture allows for geographic distribution of bookkeeping; all connections pass through a central router, which redirects requests to the correct site. The central router should be mirrored. For DC1 however, all the databases are physically at the LPSC Grenoble, and are situated on the same server.

The core packages manage the remote connection to the database, and the transmission of SQL commands. This means that we can use any database which understands SQL, and for which a java JDBC driver is available. The middle layers provide generic classes for accessing the bookkeeping databases, using their internal descriptions. Top layers of the software are provided for particular interfaces, such as the command line interface, and the web interface, or even for separate projects. The application is called the "ATLAS Metadata base Interface" (AMI).

### 6.1.2. Interfaces

In close collaboration with the production team the requirements for a "command line interface" and a "web interface" were defined.

A first version of the AMI command line interface was delivered at the end of May 2002, and an enhanced version was released at the end of July 2002.

A web interface that allows users to search the bookkeeping databases was provided later.

The development of both the command line interface and the web interface will continue, as feedback will be received from the users. The ATLAS Metadata base will be interfaced to the EU Grid WP2 package "Spitfire" [13]. This package provides a secure grid-enabled front-end to relational databases.

## 6.2.  MAGDA

Magda is being developed to fulfill the principal ATLAS '01-'02 deliverable for the Particle Physics Data Grid project [14] of a production distributed data management system deployed to users and serving BNL, CERN, and many US ATLAS grid test-bed sites.

Magda makes use of the MySQL open source relational database, Perl, Java, and C++. For data movement globftp. bbftp and scp are used. The Globus replica catalogue is currently being integrated.

The 'core' of the system is a MySQL database, but the bulk of the system is in a surrounding infrastructure for setting up and managing distributed sites with associated data locations, data store locations within those sites, and the hosts on which data-gathering servers and user applications run; gathering data from the various sorts of data stores; interfacing to users via web interfaces for presenting and querying catalog info and for modifying the system; and replicating and serving files to production and end-user applications.

All files generated for DC1 in the U.S. Grid test-bed were put in the BNL HPSS storage system using Magda. Magda also managed the replica location for these files – so more than 10000 files were automatically registered in Magda.

Magda was presented at this conference (TUCT010).

## 6.3.  Prototyping Virtual Data Approach

Because of the physics-oriented content of ATLAS Data Challenges the recipes for producing the ATLAS data (Athena jobOptions or similar "input data cards" files) have to be fully tested. The data produced have to be validated through a subsequent quality assurance and validation step. Preparation of the production recipes takes time and efforts, encapsulating considerable knowledge inside. Due to a smaller scope of ATLAS DC0 more time has been spent to assemble the proper recipes than to run the production jobs. Having the proper recipes, producing the data is straightforward. Because of the prevailing vision that the data are primary and the recipes are secondary (they needed just for the data production) it has not been clear how to treat the developed recipes after the data have been produced. It was decided to store these recipes outside of the scope of the ATLAS Bookkeeping Database AMI.

A valuable insight for ATLAS production workflow has been provided by introduction the virtual data concept. The GriPhyN [15] project provides a different perspective:

  - recipes are as valuable as the data,
  - production recipes are the virtual data.

Taking this approach to the extreme means that if you have the recipes you do not need the data (because you can reproduce them), i.e., the recipes are primary and the data are secondary. According to the virtual data architecture, recipes are stored in the virtual data catalogue database.

In the process of the ATLAS Data Challenge we have evaluated the virtual data approach for the production of several datasets. The ATLAS database group developed and delivered an infrastructure for early application of virtual data concepts and techniques to ATLAS data production. A virtual data catalogue database prototype was deployed in the spring of 2002 for evaluation in the context of the ATLAS Data Challenges. The prototype is being used successfully for data challenge event generation and detector simulation. Production job options for physics event generation and production scripts for detector simulation were recast as parameterized transformations to be catalogued, with the resulting parameterizations represented as derivations. ATLAS DC0 and DC1 parameter settings for simulations are recorded in the virtual data catalogue database.

The production system, based on the virtual data catalogue prototype, implemented the scatter-gather data processing architecture to enable high-throughput computing. The production fault tolerance has been enhanced by the use of the independent computing agents, adoption of the pull-model for agent tasks assignment (instead of push model typically used in batch production) and by the local caching of output and input data. An interesting feature provided by this architecture is the possibility for the automatic "garbage collection" in the job planner in the following sequence: production agents pull the next derivation from the virtual data catalogue; after the data has been materialized, agents register "success" in the database; when previous invocation has not been completed within the specified timeout period, it can be invoked again.

## 7.  TOOLS

Several production and bookkeeping tools were developed and used to ease the production and the monitoring of the DC1 data production.  Among them:
  - AtCom (ATLAS Commander) that allows defining submit and monitor large quantities of jobs. That tool is presented at that conference (MONT002).
  -   GRAT, the Grid Application Tools developed in the context of the US Grid projects.

These tools use several of the components described above as AMI, Magda and the relational database MySQL.

## 8.  SOFTWARE DISTRIBUTION

The ATLAS software source code is maintained at CERN in a CVS repository and then installed and compiled in a public AFS directory, under the ATLAS tree. The compilation process is done on Linux machines running CERN Red Hat Linux. Users with a good network connection and access to AFS may use executables and the data files directly linking them from CERN. This approach is anyway not indicated for remote sites with a bad connection to CERN or without access to AFS. For this purpose, a set of RPM packages has been produced, in order to install the full ATLAS software distribution on machines both with and without AFS.

Each ATLAS software release is packaged into RPM format. The kit, along with the installation script, is downloadable [16] via secure web connection.

The general criteria, followed during the package architecture development phase, have been to build a self-consistent, LINUX release independent distribution. To fulfil these requirements the RPMs have been designed to keep the same directory structure as in the CERN repository and to include the reference gcc compiler (gcc v2.95.2), the ROOT version used for the build of the release and the required libraries not part of the ATLAS software. To be consistent with the reference software, produced at CERN, the executables and libraries included in the kit are the exact copies of the files stored in the public AFS software repository.

The packages are organized in a set of base tools, required for all the installations, and several additional components. A minimal installation should provide at least the following items:
   - the set-up and management scripts;
   - the official ATLAS compilers;
   - the ROOT version using during the compilation phase;
   - the required libraries not part of the ATLAS software (external packages).
This corresponds to the ATLAS-conf, ATLAS-tools, ATLAS-release, ATLAS-compilers, ATLAS-root and ATLAS-external RPMs. If the system compiler is a gcc v2.95.2 user may choose not to install the ATLAS-compiler package. Other packages are anyway required to generate, simulate and reconstruct the data; therefore it is highly recommended that the full set of RPMs be installed on each machine.

The kit installs itself under the directory /opt/ATLAS, using about 1 GB of disk space. Relocation is also possible, providing that the change of the root directory of the kit, from /opt/ATLAS to some other place, is also reflected in the configuration scripts, by editing them after the installation. For convenience, a relocation script is included in the kit, under the /opt/atlas/etc directory.

To work with this kit, at first users must configure the environment via a set-up script (/opt/atlas/etc/atlas.shrc). After this is done, the applications are ready to be executed correctly. Some examples on how to run a simulation job are included in the kit in the ATLAS-DC1 package.

The RPM suite has proven to be robust and efficient during the first part of DC1.

Most of the countries and sites have installed the software using the official set of RPMs, but other types of installations have also been used in some sites. In particular a first one based on a full mirroring of the distributions, directly from the CERN AFS repository, and a second one from a different set of RPMs, developed by the Nordic Countries and used within the NorduGrid test-bed.

## 9.  RESOURCES

In DC1 phase1 the data needed for the HLT TDR were generated (i.e. 4-vector production using Pythia), followed by full simulation, after some selection, of the ATLAS detector response using Atlsim (Dice, Geant3). Due to the huge amount of computing time needed it was essential to make use of the computing resources available in different ATLAS institutes.

### 9.1.  Countries participating

The following 39 institutes in 18 countries participated in DC1 phase 1:
   Australia (Melbourne)
   Austria (Innsbruck)
   Canada (Alberta, CERN)
   CERN
   Czech Republic (Prague)
   France (Grenoble + Marseille; using Lyon)
   Germany (München; using FZK)
   Israel (Weizmann)
   Italy (CNAF Bologna, Frascati, Milano, Napoli, Roma)
   Japan (Tokyo)
   NorduGrid : Denmark, Norway, Sweden (Bergen, Grendel, Ingvar, ISV, LSCF, Lund, NBI, Oslo)
   Russia (Dubna, ITEP Moscow, MSU Moscow, Protvino)
   Spain (Valencia)
   Taiwan (Taipei)
   UK (Cambridge,Glasgow, Lancaster, Liverpool, RAL)
   USA (Arlington, BNL, LBNL, Oklahoma)

New countries, China, Greece, Poland and new institutes from Canada, Italy, UK and USA have joined the effort in the course of the second phase of DC1 so, now 56 institutes in 21 countries are participating to DC1. Netherlands, Romania and Switzerland intend to join the effort soon.

## 9.2.   Resources available for DC1 phase 1

In what follows we use as unit the Normalized CERN Unit (NCU) unless it is explicitly specified: 1 NCU corresponds to 1 Pentium III 500 MHz equivalent to 21 SpecInt95 (SI95).

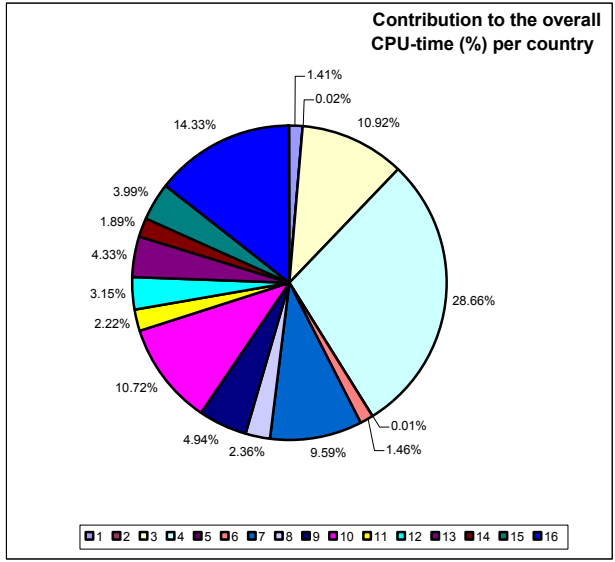

The numbers of processors per site varied between 9 and 900. At peak time we used worldwide ~3200 processors (~5000 NCUs) in 39 institutes located in 18 countries. This corresponds to ~110 kSI95 or 50% of the CPU power estimated for one Regional Facility at the LHC start-up (2007). The hardware investment made by those institutes in the last 12 months corresponds roughly to 50 % of the yearly hardware investment needed from 2006 onwards for the non-CERN part of the ATLAS Offline Computing.

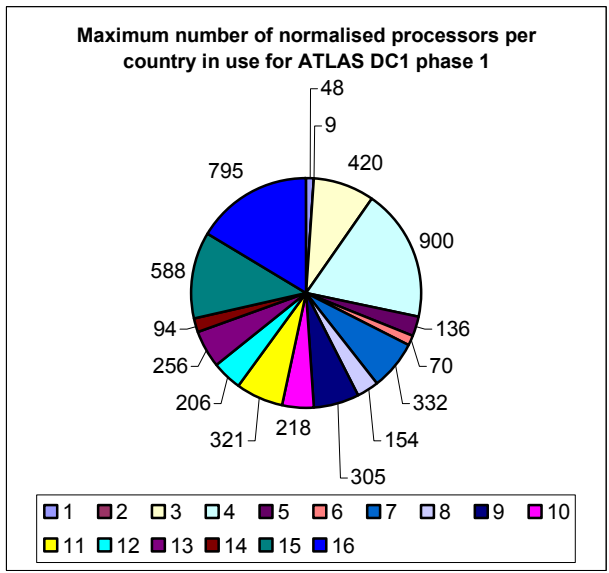

## 9.3.   Data Samples

During Phase 1 of DC1, about 50 million events in total were generated via Pythia; about 50 million events in total were passed through detailed detector simulation via Atlsim. About 40 million of the latter were single-particle events (muons, photons, electrons, pions), the remaining about 11 million ones complete physics events. This exercise took 74000 CPU-days and produced a total data volume of about 32 TBytes in about 35000 partitions. The typical characteristics are as follows:

| Event type | Output size/event (MB) | CPU-time/event (SI95sec) |
| --- | --- | --- |
| Single Particle | 0.05 | 300 |
| Minbias | 1.00 | 4000 |
| Di-jet event | 2.40 | 13000 |

The production requests from the HLT community were organized into three main parts: validation samples (very high priority), high-statistics samples (mostly high priority), and medium-statistics samples (ranging from low to high priority). A monitoring web working page [17] was set up to monitor the progress of the production activities. In addition to the original information (physics contents, simulation specifications, number of events, priority), this page contains also organizational (dataset numbers, groups-in-charge, status, etc.) and statistical (numbers of generated/simulated events, time to process one/all events, etc.) information relevant for the individual sub-samples. Most of this information is also accessible from the bookkeeping database. The data at CERN are stored in the CASTOR system.

The samples assigned the highest priority were the validation samples [18]. They consist of single-particle events, jet-scan samples, and some physics event channels taken from old TDR tapes. About 740k events were processed and 110 GBytes of data were produced, the time needed being about 900 CPU days.

The most challenging part, with respect to CPU and data storage requirements, was the production of the high-statistics samples [19]. They consist of 36 million single-muon events, about 5 million di-jet events of different $E_T$(hard scattering) cuts (applying particle-level filtering or not), and 1 million minimum-bias events simulated with different $|\eta|$ cuts. The data volume of the whole sample amounts to about 15 TBytes, the total CPU time needed to about 44000 days. Note that not all the produced data are stored in the CERN CASTOR system about 10 TBytes is kept at different production sites.

The medium-high statistics samples [20] comprise production requests by various subgroups of the HLT community: the e/gamma, Level-1, jet/ETmiss, B-physics, b-jet, and muon trigger groups; and a sample of about 80k single pions. The e/gamma samples contain a huge production of single-electron (1.1 million) and single-photon (1.6 million) events at different energies and η values. Sub-samples of the B-physics trigger and b-jet trigger samples were simulated for the Inner Detector only, the rest either with the "central" detector (ID+Calorimeters; e.g., the e/gamma single-particle production) or the full detector. All the generated events for the B-physics trigger group were taken from existing TDR tapes. All the about 7 million simulated events correspond to a data volume of about 9 TBytes, the total CPU time necessary to process them was 30000 days.

In summary, the total estimated data volume produced during DC1/1 is about 24 TBytes and about 8 TBytes for generated events; the total CPU time necessary to generate all the events was about 1000 days, the time to simulate all the events about 74000 days.

## 10. DC1 AND THE GRID

Recent advances in computing can be characterized by emerging Grid technologies. Powered by various middleware, Grid computing infrastructures are becoming a reality, and as such are particularly important for large distributed projects like the High Energy Physics experiments, and ATLAS in particular. By harnessing distributed and scarce resources into a powerful system, the Grid is expected to play a major role in a not-so-distant future. Apart of optimization of the distributed resources usage, the Grid will naturally offer all the collaboration members a uniform way of carrying out computing tasks. This is essential for large production tasks, which need plenty of resources, both hardware and human, worldwide.

Data Challenges are the perfect opportunity to evaluate the current status of the Grid middleware and assess what has to be done by the collaboration in order to make a smooth transition to the Grid tools. Therefore ATLAS has been extremely active in Grid matters over the last few months.

A significant fraction of the first phase of the ATLAS DC1 was performed in the Grid environment, involving 11 out of 39 sites.

All the production of the Nordic countries was processed on NorduGrid [21]. The test-bed included 8 Linux clusters across the Scandinavia. Despite having different operating systems and hardware characteristics, the clusters performed as a single farm, having jobs distributed in optimal way, and writing the output onto a dedicated storage area at the Oslo University. A detailed report can be found on the web [22] and a presentation has been given in this conference (MOCT011).

Three sites of the US Grid test-bed took part in DC1 phase 1. About 10% of the US contribution to this phase was done on this test-bed.

In phase 2 again all the Nordic production was done on NorduGrid.
On the US side for the pile-up exercise 6000 jobs have been run that corresponds to 8 CPU years and 10 TB of data being used (input and output).
For both the test-beds have been intensively used for the reconstruction step.

On the European side an ATLAS-EDG task force was put in place in August 2002. A first test was successfully run in September 2002 followed by other tests at different periods. It is intended to run a fraction of the reconstruction in the next coming weeks on the EDG test-bed. During that summer the ATLAS-EDG task force will start to work on the LCG-1 prototype.

## 11. CONCLUSIONS

ATLAS Data Challenge 1 ran from spring 2002 to spring 2003. Due to several constraints it was divided in several phases.

Phase 1 was used to put in place the worldwide production infrastructures and to produce the bulk of simulated data needed by our colleagues of the High Level Trigger who have to prepare a Technical Design Report. Over a period of 40 calendar days the equivalent of 1.5 million of SI95-days were used to produce 10 million of physics events and 40 million of single particle events for a total volume of 30 TBytes. That volume was certainly over our more optimistic expectations. 39 institutes in 18 countries actively participated to the effort.

The pile-up production in the second phase ran smoothly. 135000 SI95-days were necessary to produce about 20 TBytes of data.

Most of that data has already been reconstructed. A new reconstruction round will be needed with in addition to the standard 'off-line' reconstruction algorithms the Level 1 and Level 2 specific trigger ones. This could probably be done in about 1 month.

During that exercise we have seen the emergence of the production on the Grid. Grid tools were used intensively on NorduGrid and US test-beds. We are confident that their use will continue to grow.

Finally may be the most important benefits of DC1 have been to establish a very good collaborative effort between all members of the DC team and to increase the momentum of the ATLAS computing as a whole.

## Acknowledgments

We would like to thank all members of the ATLAS collaboration who participated to the effort.

It would not have been possible to successfully run our Data Challenge without the involvement of numerous people from each institutes and computing centers who helped us to put in place the production chain and to run the production. We, sincerely, thank all of them.

## References

[1] http://lhc-computing-review-public.web.cern.ch/lhc-computing-review-public/Public/Report_final.PDF

[2] http://www.cern.ch/Atlas

[3] http://lcg.web.cern.ch/lcg/

[4] CERN Program Library Q100/Q101

[5] http://root.cern.ch/

[6] http://mdobbs.web.cern.ch/mdobbs/HepMC/

[7] http://paw.web.cern.ch/paw/

[8] http://castor.web.cern.ch/castor/

[9] http://wwwasdoc.web.cern.ch/wwwasdoc/shortwrupsdir/i202/top.html

[10] http://atlas.web.cern.ch/Atlas/GROUPS/SOFTWARE/DC/Validation/www/

[11] http://larbookkeeping.in2p3.fr/AMI/

[12] http://www.atlasgrid.bnl.gov/magda/info

[13] http://hep-proj-spitfire.web.cern.ch/hep-proj-spitfire/server/doc/index.html

[14] PPDG project

[15] http://www.griphyn.org

[16] https://classis01.roma1.infn.it/atlas-farm/atlas-kit

[17] http://atlasinfo.cern.ch/ATLAS/GROUPS/SOFTWARE/DC/DC1/DC1_1/production_requests.html

[18] http://atlasinfo.cern.ch/ATLAS/GROUPS/SOFTWARE/DC/DC1/DC1_1/validation/validation_samples.html

[19] http://atlasinfo.cern.ch/ATLAS/GROUPS/SOFTWARE/DC/DC1/DC1_1/highStat/high_statistics_samples.html

[20] http://atlasinfo.cern.ch/ATLAS/GROUPS/SOFTWARE/DC/DC1/DC1_1/mediumStat/medium_statistics_samples.html

[21] http://www.nordugrid.org

[22] http://www.nordugrid.org/documens/ATLASdc1.html